\documentstyle[12pt]{article}

\renewcommand{\thesection}{\Roman{section}. }

\def\overlay#1#2{\ifmmode%
\setbox0=\hbox{$#1$}%
\setbox1=\hbox to\wd0{\hss$#2$\hss}\else%
\setbox0=\hbox{#1}%
\setbox1=\hbox to\wd0{\hss#2\hss}\fi%
 #1\hskip-\wd0\box1 }

\def\sumint{\hbox{$\sum$}\!\!\!\!\!\!\int}

\begin{document}
\hfill\vbox{
\hbox{NUHEP-TH-95-2}
\hbox{hep-ph/9501375}
\hbox{January 1995} }\par
\thispagestyle{empty}

\begin{center}
{\Large \bf Effective Field Theory Approach to High-Temperature Thermodynamics}

\vspace{0.15in}

Eric Braaten and Agustin Nieto\\

{\it Department of Physics and Astronomy, Northwestern University,
Evanston, IL 60208}

\end{center}

\begin{abstract}
An effective field theory approach is developed for calculating the
thermodynamic properties of a field theory at high temperature $T$
and weak coupling $g$.  The effective theory is the 3-dimensional
field theory obtained by dimensional reduction to the bosonic
zero-frequency modes.
The parameters of the effective theory can be calculated as perturbation
series in the running coupling constant $g^2(T)$.  The free energy is
separated into the contributions from the momentum
scales $T$ and $gT$, respectively.  The first term can be written as
a perturbation series in $g^2(T)$. If all forces are screened at the scale
$gT$, the second term can be calculated as a perturbation series in $g(T)$
beginning at order $g^3$.  The parameters of the effective theory
satisfy renormalization group equations that can be used to sum up
leading logarithms of $T/(gT)$.  We apply this method to a massless scalar
field with a $\Phi^4$ interaction, calculating the free energy to order
$g^6 \log g$ and the screening mass to order $g^5 \log g$.

\end{abstract}

\newpage
\begin{center}\section{Introduction}\end{center}

When matter is subjected to sufficiently extreme temperature or density,
both quantum and relativistic effects become important.
Such conditions arise in astrophysics and in cosmology and it may be
possible to create them experimentally using heavy-ion collisions.
A system at such an extreme temperature or density is most appropriately
described by quantum field theory.  The thermodynamic functions that
describe the bulk equilibrium properties of such a system are given by
the free energy density and its derivatives.
If the temperature $T$ is high enough that all masses can be neglected,
the free energy density depends only on $T$ and on the
coupling constants of the field theory.

In recent years, there have been significant advances in the perturbative
calculation of the free energy for high temperature field theories.
The free energy for a massless scalar field with a $\Phi^4$ interaction
was computed to order $g^4$ by Frenkel, Saa, and Taylor~\cite{fst} in 1992,
and the order-$g^5$ correction was recently calculated
by Parwani and Singh \cite{parwani-singh}.
The free energy for high temperature QED was calculated to order $e^4$ by
Coriano and Parwani~\cite{parwani-coriano}, and extended to order $e^5$ by
Parwani~\cite{parwani2}.
The free energy for a quark-gluon plasma in the high temperature limit
was recently calculated to order $g^4$ by Arnold and Zhai~\cite{arnold-zhai}.
The importance of these calculations goes far beyond simply determining
one more term in the perturbation series.  The leading term in the
perturbation series is just the free energy of an ideal gas.  The
order-$g^2$ correction takes into account interactions between
particles in the ideal gas.  At order $g^3$, there is a qualitatively
new contribution to the free energy.  If the
temperature is large compared to the masses of the particles, the force
mediated by the exchange of a particle is long-range compared to the
typical separation of the particles, which is of order $1/T$.
The system therefore behaves like a plasma,
screening the long-range interaction beyond the scale $1/(gT)$.
It is this screening that is responsible for the $g^3$ term in the
free energy density.
Because of renormalization effects, the corrections of order $g^4$ and $g^5$
are also important.  The coupling constant $g(\mu)$ depends on an arbitrary
renormalization scale $\mu$.  The resulting ambiguity in the leading
nontrivial term in a perturbative expansion can only be reduced by a
next-to-leading order calculation.
While it may be clear on physical grounds that the scale $\mu$
should be of order $T$, the difference between the choices $g(T)$
and $g(2 \pi T)$ can be of great practical significance.
Thus a calculation to order $g^4$
is needed in order to determine the appropriate scale $\mu$
in the order-$g^2$ correction to the ideal gas term, while a calculation
to order $g^5$ is required in order to determine the scale in the
order-$g^3$ plasma term.

In the case of QCD, there is also a qualitatively new effect that arises
at order $g^6$.  As pointed out by Linde~\cite{linde,gpy} in 1979,
the loop expansion for the free energy breaks down at this order in $g$.
Up to order $g^5$, the free energy can be calculated  using
a resummation of perturbation theory that takes into account the screening
of the chromoelectric force at distances of order $1/(gT)$.
However the chromomagnetic force is not screened at the scale $gT$,
and this causes a breakdown in the resummed perturbation expansion at
order $g^6$.  This breakdown has been widely interpreted as implying
that nonperturbative effects enter in at this order, rendering the
perturbation series meaningless beyond order $g^5$.
This longstanding problem was recently solved
by constructing  a sequence of two effective field
theories that are equivalent to thermal QCD over successively longer
length scales~\cite{braaten}.  The first effective theory reproduces the
static gauge-invariant correlators of thermal QCD at distances of order
$1/(gT)$ or larger, while the second effective theory reproduces
the correlators at distances of order $1/(g^2T)$ or
larger.  Using this construction, the free energy can be separated into
contributions from the momentum scales $T$, $gT$, and $g^2T$,
with well-defined weak coupling expansions that begin at order
$g^0$, $g^3$, and $g^6$, respectively.  The contributions from the scales
$T$ and $gT$ can be computed using perturbative methods.
For the contribution from the scale $g^2T$, the coefficients in the
weak-coupling expansion can be calculated
using lattice simulations of pure-gauge QCD in 3 Euclidean dimensions.
This result demonstrates the power of the effective-field-theory approach.
This power can also be brought to bear on other fundamental problems
in thermal field theory.  For example, it has also been used to
determine the correct asymptotic behavior of the correlator of Polyakov
loop operators in high temperature QCD~\cite{braaten-nieto}.

Effective field theory provides an effective method for unravelling the
effects of the various momentum scales that arise in field theory at
high temperature.  In addition to providing insight into the qualitative
behavior of the theory, effective field theory can also be used to streamline
perturbative calculations, such as those in
Refs.~\cite{fst}-\cite{arnold-zhai}.  In this paper, we use
effective field theory to develop a practical method for
calculating the thermodynamic functions of a field theory at
high temperature $T$ and weak coupling $g$.  To illustrate the method,
we apply it to a massless scalar field theory with a $\Phi^4$
interaction.  In section II, we explain how the effective field theory
that describes distance scales of order $1/(gT)$ or larger is
related to dimensional reduction to the bosonic zero-frequency modes.
In Section III,  we show how the parameters of the effective theory
can be obtained by matching perturbative calculations in the effective
theory and in the full theory.  In Section IV, we use the effective theory
for the massless $\Phi^4$ theory to calculate the free energy
to order $g^5$ and the screening mass to order $g^4$.
The accuracy of these calculations is improved in Section V to order
$g^6 \log g$ for the free energy and to order $g^5 \log g$ for the
screening mass by using renormalization group equations for the
parameters of the effective theory.
We conclude in Section VI with a discussion of the application of our
effective-field-theory approach to gauge theories and with a brief
comparison with related work.  In appendices A and B, we collect all the
necessary formulas for sum-integrals in the full theory and for integrals
in the 3-dimensional effective theory.  In appendix C, we derive the
renormalization group equations for the parameters of the effective theory.

\begin{center} \section{Dimensional Reduction} \end{center}

In the limit of high temperature $T$, the static correlation functions
of a field theory in (3+1) dimensions can be reproduced at long distances
$R \gg 1/T$ by an effective field theory in 3 dimensions.
This idea, which is called ``dimensional reduction'',
has a long history \cite{gpy,appelquist-pisarski,nadkarni,landsman}.
It has provided insight
into the qualitative behavior of the field theory at high temperature,
but it has never been fully exploited for quantitative calculations.

Dimensional reduction is based on the fact that static correlation
functions for a field theory in thermal equilibrium
can be expressed in terms of Euclidean functional integrals.
The partition function ${\cal Z}$ is defined by
\begin{equation}
{\cal Z}(T) \;=\; {\rm trace} \left( e^{- \beta H} \right) ,
\label{Zdef}
\end{equation}
where $H$ is the hamiltonian operator and $\beta = 1/T$.
The operator $e^{- \beta H}$ is the evolution operator that evolves
a state from time $t=0$ to the imaginary time $t=-i \beta$.
It can be represented as a functional integral over fields $\Phi({\bf x},t)$
defined on the time interval from 0 to $-i \beta$. It is natural to
change variables from $t$ to the imaginary time $\tau = i t$.
The partition function (\ref{Zdef}) is then given by the
Euclidean functional integral
\begin{equation}
{\cal Z}(T) \;=\; \int {\cal D} \Phi({\bf x},\tau) \; \exp
\left( - \int_0^\beta d \tau \int d^3x \; {\cal L} \right) ,
\label{Z}
\end{equation}
where ${\cal L}$ is the negative of the lagrangian density for the
(3+1)-dimensional theory with the time $t$ analytically continued to
$- i \tau$.  The trace in (\ref{Zdef}) is implemented by imposing
boundary conditions on the fields:
\begin{equation}
\Phi({\bf x},\tau=\beta) \;=\; \pm \; \Phi({\bf x},\tau=0),
\label{bc}
\end{equation}
where the plus sign holds for bosonic fields and the minus sign for
fermionic fields.
The correlator of two operators ${\cal O}({\bf 0})$ and
${\cal O}({\bf R})$ is obtained by averaging their product over fields
$\Phi({\bf x},\tau)$ with the exponential weighting factor in (\ref{Z}).

Because of the periodicity conditions (\ref{bc}) on the fields, they can be
decomposed into Fourier modes in $\tau$, with Matsubara frequencies
$\omega_n = 2n \pi T$ for bosons and $\omega_n = (2n+1)\pi T$ for fermions.
The contribution to a correlator from the exchange of a Fourier mode
with frequency $\omega_n$ falls off at large $R$
like $\exp(-|\omega_n| R)$.  Thus the only modes whose
contributions do not fall off exponentially at distances greater
than $1/T$ are the $n=0$ modes of the bosons.  This suggests the strategy
of integrating  out the fermionic modes and the nonzero modes of the bosons
to get an effective theory for the bosonic zero modes.  This process is
called ``dimensional reduction''.  It results in a 3-dimensional
Euclidean field theory with bosonic fields only which reproduces the static
correlators of the original theory at distances $R \gg 1/T$.

Constructing the dimensionally-reduced effective theory by
actually integrating out degrees of freedom is cumbersome beyond leading
order in the coupling constant.  Once the appropriate 3-dimensional fields
and their symmetries have been identified, a better strategy is to use the
methods of ``effective field theory'' \cite{lepage}.  One writes down the most
general lagrangian ${\cal L}_{\rm eff}$ for the 3-dimensional fields that
respects the symmetries.  This effective lagrangian has infinitely many
parameters, but they are not arbitrary.  By computing static correlators
in the full theory, computing the corresponding correlators
in the effective theory, and demanding that they agree at distances
$R \gg 1/T$, one can determine the parameters of ${\cal L}_{\rm eff}$
in terms of $T$ and the parameters of the original theory.
Note that this matching procedure does not necessarily require the
explicit determination of the relation between the fields in the effective
theory and the fundamental fields.

The construction of the 3-dimensional effective theory is complicated
by ultraviolet divergences.  The ultraviolet divergences associated
with the original 4-dimensional theory are removed by the standard
renormalization procedure.  However the 3-dimensional effective theory
also has ultraviolet divergences whose origin can be traced
to integrating out the nonzero-frequency modes in the full theory.
They must be regularized by introducing an ultraviolet cutoff $\Lambda$.
The parameters in ${\cal L}_{\rm eff}$ must therefore depend on $\Lambda$
in such a way as to cancel the $\Lambda$-dependence of the regularized loop
integrals in the effective theory.
The natural scale for the ultraviolet cutoff $\Lambda$ is of order $T$,
since this is the scale at which the corresponding integrals are cut off
by the nonzero modes in the full theory.
It is useful, however, to keep the cutoff $\Lambda$ independent of $T$,
so that the only dependence on $T$ in the effective theory comes
from the parameters in ${\cal L}_{\rm eff}$.
{}From the point of view of the full theory, the ultraviolet cutoff $\Lambda$
of the effective theory plays the role of an arbitrary factorization scale
that is introduced to separate the momentum scale $T$ from lower
momentum scales, such as $gT$, which can be described within the effective
theory.

The ultraviolet divergences of the effective theory include power
ultraviolet divergences of the form $\Lambda^p$, $p=1,2,3, \ldots$,
and logarithmic ultraviolet divergences of the form $\log(\Lambda/m)$,
where $m$ is a mass scale in the effective theory.  The power divergences
and the logarithmic divergences are quite different in character.
The coefficients of the power divergences depend on the regularization
procedure and are therefore simply regularization artifacts.
In particular, the coefficient of $\Lambda^p$ in some observable
calculated in the effective theory is independent of the coefficient
of $T^p$ for that observable in the full theory.
Therefore power divergences of the form $\Lambda^p$ from loop integrals
must be completely cancelled by terms proportional to $\Lambda^p$
in the parameters of ${\cal L}_{\rm eff}$.  In contrast to the power
divergences, logarithmic ultraviolet divergences have coefficients
that are independent of the regularization procedure.  The reason for
this is that a logarithmic ultraviolet divergence of the form
$\log(\Lambda/m)$ from a loop integral must match onto a
$\log(T/\Lambda)$ term in one of the parameters of ${\cal L}_{\rm eff}$
in order for the $\Lambda$-dependence to cancel.  Thus the logarithmic
divergences in the effective theory are related to logarithms of $T$
in the full theory and therefore have real physical significance.
Because of the unphysical character of power ultraviolet divergences,
it is convenient to use a regularization procedure for the effective
theory in which power ultraviolet divergences are subtracted and the
remaining logarithmically divergent integrals are cut off at the scale
$\Lambda$.  With such a regularization procedure, only the logarithmic
divergences need to be cancelled by the parameters of the effective theory.
The subtraction of power divergences can be implemented with any cutoff
procedure, including a momentum cutoff or lattice regularization.
For perturbative calculations, a particularly convenient cutoff procedure
is dimensional regularization, in which momentum integrals are analytically
continued to $3-2 \epsilon$ spacial dimensions.
Power divergences are automatically subtracted with this method,
because integrals without any momentum scale are 0 by definition
in dimensional regularization.  The remaining logarithmic
ultraviolet divergences appear as poles in $\epsilon$, and the
renormalization procedure can be completed by subtracting these poles.
If $\Lambda$ is the momentum scale introduced by dimensional
regularization, then this ``minimal subtraction'' procedure is equivalent
to cutting off the logarithmic ultraviolet divergences
at a momentum scale of order $\Lambda$.

We turn now to the specific case of a massless scalar field with
a $\Phi^4$ interaction.  In the partition function (\ref{Z}),
the Euclidean lagrangian density is
\begin{equation}
{\cal L} \;=\;
{1 \over 2} \left( \partial_\tau \Phi \right)^2
\;+\; {1 \over 2} \left( \makebox{\boldmath $\nabla$} \Phi \right)^2
\;+\; {1 \over 4!} g^2 \Phi^4 .
\label{L}
\end{equation}
The 3-dimensional effective field theory obtained by dimensional
reduction describes a scalar field $\phi({\bf x})$
that can be approximately identified with the zero-frequency mode
of the field in the original theory:
\begin{equation}
\sqrt{T} \int_0^\beta d \tau \; \Phi({\bf x},\tau) \;\approx\; \phi({\bf x}) .
\label{phi}
\end{equation}
The symmetries of the effective theory are rotational symmetry and the
discrete symmetry $\phi({\bf x}) \to - \phi({\bf x})$, which follows from
the symmetry $\Phi \to - \Phi$ of the fundamental lagrangian (\ref{L}).
The effective lagrangian has the general form
\begin{equation}
{\cal L}_{\rm eff} \;=\;
{1 \over 2} \left( \makebox{\boldmath $\nabla$} \phi \right)^2
\;+\; {1 \over 2} m^2(\Lambda) \; \phi^2
\;+\; {1 \over 4!} \lambda(\Lambda) \; \phi^4
\;+\; \delta {\cal L},
\label{Leff}
\end{equation}
where $\delta {\cal L}$ includes all other local terms that are consistent
with the symmetries.  The parameters $m^2(\Lambda)$ and $\lambda(\Lambda)$
and the infinitely many parameters in (\ref{Leff}) depend on
the ultraviolet cutoff $\Lambda$, the temperature $T$, and the
coupling constant $g^2$.
The partition function (\ref{Z}) can be expressed as a Euclidean functional
integral over the 3-dimensional field $\phi({\bf x})$:
\begin{equation}
{\cal Z}(T) \;=\;
e^{- f(\Lambda) \, V} \int^{(\Lambda)} {\cal D} \phi({\bf x})
\exp \left( - \int d^3x \; {\cal L}_{\rm eff} \right) .
\label{Zeff}
\end{equation}
The parameter $f(\Lambda)$ in the exponential prefactor is the coefficient
of the unit operator in the effective lagrangian, which was omitted from
${\cal L}_{\rm eff}$ in (\ref{Leff}).  In addition to depending on
$g^2$ and $T$, it also depends on $\Lambda$
in such a way as to cancel the $\Lambda$-dependence of the
functional integral in (\ref{Zeff}).
The correlator of operators ${\cal O}({\bf 0})$ and
${\cal O}({\bf R})$ in the full theory
can be calculated at long distances $R \gg 1/T$
by identifying the corresponding operators in the effective theory
and averaging their product over fields $\phi({\bf x})$ with the exponential
weighting factor in (\ref{Zeff}).

Renormalization theory implies that correlators at long distances
$R \gg 1/T$ can be reproduced to any desired accuracy by adding
sufficiently many operators to the effective lagrangian and tuning their
coefficients with sufficient accuracy as functions of $g^2$, $T$, and
$\Lambda$ \cite{lepage}.  With the three terms in the
lagrangian that are given explicitly in (\ref{Leff}),
long-distance correlators can only be reproduced with limited accuracy.
If we include the operator $\phi^6$, the field theory becomes
renormalizable, but this renormalizable theory is still only accurate up to
a finite order in the coupling constant $g$.  This has been interpreted
as a breakdown of dimensional reduction \cite{landsman}, but
the correct interpretation is simply that nonrenormalizable
operators must also be included in ${\cal L}_{\rm eff}$ in order to extend the
accuracy to higher order in $g$.  In general, one must include all
operators that are invariant under rotations and under the symmetry
$\phi \to - \phi$.  The resulting field theory is nonrenormalizable
and has infinitely many parameters.  These parameters, however, are not
arbitrary, but are determined as functions of $g^2$, $T$,
and $\Lambda$ by the condition that the effective theory
reproduce the long distance behavior of the original theory.

It is easy to determine the magnitude of the coefficient of a
general operator in the effective lagrangian.  From the kinetic term
$(\makebox{\boldmath $\nabla$} \phi)^2$ in (\ref{Leff}), we see that
the field $\phi$ should be assigned a scaling dimension of $1/2$.
The operators given explicitly in (\ref{Leff}) then have dimensions
3, 1, and 2.  If we use a renormalization procedure in which power
ultraviolet divergences are subtracted, then
by dimensional analysis, an operator of dimension $d$ in the effective
lagrangian must have a coefficient that is proportional to $T^{3-d}$.
It remains only to determine its order in $g$.
The operator $\phi^4$ is generated at tree level from the $g^2 \Phi^4$
term in the original lagrangian, and therefore has a coefficient
proportional to $g^2$.  All other interaction terms in ${\cal L}_{\rm eff}$
arise from loop diagrams in the effective theory.  Operators with
$2n$ powers of $\phi$ are generated by 1-loop diagrams with $n$ 4-point
interactions and therefore have coefficients of order $g^{2n}$.
Thus an operator with the schematic structure
$\nabla^{2m} \phi^{2n}$ has dimension $d = 2m+n$ and will appear in
${\cal L}_{\rm eff}$ with a coefficient of magnitude $g^{2n} T^{3-d}$.
The case $n=1$ is an exception, because the 1-loop diagram with
two external lines is momentum-independent and therefore only the operator
$\phi^2$ is generated at order $g^2$.  Operators of dimension $d = 2m+1$
with the schematic structure $\nabla^{2m} \phi^2$ have coefficients
with magnitude $g^4 T^{3-d}$ for $m \ge 2$.
The only other exception is $\phi^4$, which is generated
at tree level and has a coefficient of magnitude $g^2T$.

\begin{center} \section{Short-distance Coefficients} \end{center}

The coefficients of the operators in the effective lagrangian
(\ref{Leff}) must be tuned
as functions of $g$, $T$, and $\Lambda$ so that the effective theory
reproduces the static correlation functions of the full theory at distances
$R \gg 1/T$.  The parameters can be determined by computing various
static quantities in the full theory, computing the corresponding
quantities in the effective theory, and demanding that they match.
If the running coupling constant $g^2(T)$ of the full theory is small
at the scale $T$, then it is convenient to carry out these calculation
using perturbation theory in $g^2(T)$.
A strict perturbation expansion in $g^2$ is afflicted with infrared
divergences due to long-range forces mediated by massless particles.
These divergences are screened at the scale $gT$, but this screening can
only be taken into account by summing up infinite sets of higher-order
diagrams.
This breakdown of perturbation theory does not prevent
its use as a device for determining the
short-distance coefficients in the effective lagrangian.
As long as we can carry out perturbative calculations
in the effective theory that make the same incorrect assumptions about the
long-distance behavior as perturbation theory in the full theory,
we can match the results and determine the short-distance coefficients.

In the case of the $\Phi^4$ theory, conventional perturbation theory in $g^2$
corresponds to decomposing the lagrangian (\ref{L}) as
${\cal L} = {\cal L}_{\rm free} + {\cal L}_{\rm int}$, where
\begin{eqnarray}
{\cal L}_{\rm free}
&=& {1 \over 2} (\partial_\tau \Phi)^2
	\;+\; {1 \over 2} (\makebox{\boldmath $\nabla$} \Phi)^2,
\nonumber \\
{\cal L}_{\rm int} &=& {g^2 \over 4!} \Phi^4 .
\label{Lpert}
\end{eqnarray}
We will refer to the resulting perturbation theory as a ``strict''
perturbation expansion in $g^2$.
The free part of the lagrangian describes a massless scalar field.
A mass will not be generated at any finite order in $g^2$,
and the absence of a mass will give rise to infrared divergences
that become more and more severe as you go to higher and higher orders
in $g^2$.  This behavior is physically incorrect.  It will be clear from
the effective theory that a mass $m$ of order $gT$ is generated
by higher loop diagrams and it provides the screening that
cuts off the infrared divergences.  One way of dealing with
the infrared divergences in the full theory is to use a reorganization
of perturbation theory that incorporates the effects of the mass $m$
in the free part of the lagrangian.  The simplest possibility \cite{parwani1}
is to write the lagrangian (\ref{L})
as ${\cal L} = {\cal L}_{\rm free} + {\cal L}_{\rm int}$, where
\begin{eqnarray}
{\cal L}_{\rm free}
&=& {1 \over 2} (\partial_\tau \Phi)^2
	\;+\; {1 \over 2} (\makebox{\boldmath $\nabla$} \Phi)^2
	\;+\; {1 \over 2} m^2 \Phi^2 ,
\nonumber \\
{\cal L}_{\rm int}
&=& {g^2 \over 4!} \Phi^4 \;-\; {1 \over 2} m^2 \Phi^2 .
\end{eqnarray}
Both $g^2$ and $m^2$ in the interaction term are treated as perturbation
parameters of the same order.  The mass parameter $m^2$ must be of order
$g^2 T^2$ in order to avoid large perturbative corrections that grow
quadratically with $T$ in the high temperature limit.  Another possibility
is to add the mass term to ${\cal L}_{\rm free}$ and subtract it from
${\cal L}_{\rm int}$ only for the zero-frequency mode of
$\Phi({\bf x},\tau)$ \cite{arnold-espinoza}.
Both of these approaches have the drawback that the resulting sums
and integrals involve two momentum scales $T$ and $m$, making calculations
unnecessarily difficult.

A simpler approach is to calculate in both the full theory
and the effective theory using
ordinary perturbation theory in $g^2$, but with an infrared cutoff to
regularize the infrared divergences.  In the full theory, this strict
perturbation expansion in $g^2$ is defined by the decomposition (\ref{Lpert}).
In the effective lagrangian (\ref{Leff}), the coefficients $m^2$ and
$\lambda$ are of order $g^2$ and all the coefficients in $\delta {\cal L}$
are of order $g^4$ or higher.  Thus, in the effective theory,
the strict expansion in $g^2$ is defined by the decomposition
${\cal L}_{\rm eff} = \left({\cal L}_{\rm eff} \right)_{\rm free}
+ \left({\cal L}_{\rm eff} \right)_{\rm int}$, where
\begin{eqnarray}
\left({\cal L}_{\rm eff} \right)_{\rm free}
&=& {1 \over 2} \left( \makebox{\boldmath $\nabla$} \phi \right)^2 ,
\nonumber \\
\left({\cal L}_{\rm eff} \right)_{\rm int}
&=& {1 \over 2} m^2(\Lambda) \; \phi^2
\;+\; {1 \over 4!} \lambda(\Lambda) \; \phi^4
\;+\; \delta {\cal L}.
\label{Leffpert1}
\end{eqnarray}
The expansions in $g^2$ defined by (\ref{Lpert}) and (\ref{Leffpert1})
both generate infrared divergences
that become more and more severe in higher orders.
But if the parameters in the effective lagrangian are tuned
so that the two theories are equivalent at long distances,
then the infrared divergences in their strict perturbative expansions
will also match.  Thus, in spite of the fact that the strict expansion
in $g^2$ gives a physically incorrect treatment of infrared effects,
we can use it as a device for computing short-distance coefficients.

\subsection{Coefficient of the unit operator}

In this subsection, we calculate the parameter $f$ in (\ref{Zeff})
to next-to-next-to-leading order in $g^2$.  The parameter $f$ can be
interpreted as the coefficient of the unit operator which has been
omitted from the effective lagrangian (\ref{Leff}).
We will determine $f$ by matching calculations of $\log{\cal Z}$
in the full theory and in the effective theory.

We first calculate $\log {\cal Z}$ to next-to-next-to-leading order in $g^2$
using the perturbation expansion for the full theory defined
by the decomposition (\ref{Lpert}).
It is given by the sum of the Feynman diagrams in Fig.~1:
\begin{eqnarray}
{T \; \log {\cal Z} \over V}
&\approx& - {1 \over 2}\; \sumint_P \log(P^2)
\;-\; {Z_g^2 g^2 \over 8} \left( \sumint_P {1 \over P^2} \right)^2
\nonumber \\
&& \;+\; {g^4 \over 16} \left( \sumint_P {1 \over P^2} \right)^2
	\sumint_P {1 \over (P^2)^2}
\;+\; {g^4 \over 48} \; \sumint_{PQR} {1 \over P^2 Q^2 R^2 (P+Q+R)^2} ,
\label{logZint}
\end{eqnarray}
where the sum-integral notation is defined in Appendix~A.
We regularize both ultraviolet and infrared divergences using dimensional
regularization in $3 - 2 \epsilon$ spatial dimensions, taking the
momentum scale introduced by dimensional regularization to be $\Lambda$.
In (\ref{logZint}) and below,
we use the symbol ``$\approx$'' for an equality that
holds only in a strict perturbation expansion in powers of $g^2$.
Such an equality does not properly take into account the screening of
infrared divergences at the scale $gT$, but it can be used for determining
short-distance coefficients.
The sum-integrals appearing in (\ref{logZint}) are given in Appendix~A.
To the order that is required, renormalization of the coupling constant
in the $\overline{\rm MS}$ scheme is accomplished by the substitution
\begin{equation}
Z_g \;=\;
1 \;+\; {3 \over 4 \epsilon} {g^2 \over 16 \pi^2} .
\end{equation}
After this renormalization, the final result is
\begin{eqnarray}
{T \, \log {\cal Z} \over V} &\approx& {\pi^2 \over 9} T^4
\Bigg\{ {1 \over 10} \;-\; {1 \over 8} {g^2(\Lambda) \over 16 \pi^2}
\nonumber \\
&& \qquad \qquad \;+\; {1 \over 8}
\left[ 3 \log{\Lambda \over 4 \pi T} + {31 \over 15} + \gamma
+ 4 {\zeta'(-1) \over \zeta(-1)} - 2 {\zeta'(-3) \over \zeta(-3)} \right]
\left( {g^2 \over 16 \pi^2} \right)^2 \Bigg\} ,
\label{logZ}
\end{eqnarray}
where $\gamma$ is Euler's constant and $\zeta(z)$ is the Riemann zeta function.
The apparent dependence of the right side of (\ref{logZ}) on $\Lambda$
is illusory.  The renormalization group equation for the coupling constant,
\begin{equation}
\mu {d \ \over d \mu} {g^2 \over 16 \pi^2}
\;=\; 3 \left( {g^2 \over 16 \pi^2} \right)^2 \;+\; O(g^6),
\label{dg}
\end{equation}
implies that the explicit logarithmic dependence on $\Lambda$ in the $g^4$
term of (\ref{logZ}) is cancelled by the $\Lambda$-dependence of the coupling
constant in the $g^2$ term.  Thus, up to corrections of order $g^6$,
we can replace $\Lambda$ on the right side
of (\ref{logZ}) by an arbitrary renormalization scale $\mu$.

We now consider $\log {\cal Z}$ in the effective theory:
\begin{equation}
\log {\cal Z} \;=\; - f(\Lambda) \; V \;+\; \log {\cal Z}_{\rm eff}.
\label{logZeff1}
\end{equation}
The partition function ${\cal Z}_{\rm eff}$ for the effective theory
is the functional integral in (\ref{Zeff}).  In the diagrammatic expansion
for $\log {\cal Z}_{\rm eff}$, the leading terms are given by the
diagrams in Fig.~1, plus additional diagrams involving mass insertions
as in Fig.~2.  To match with the strict expansion in $g^2$
for the full theory, we should
calculate in the effective theory using the perturbation theory
defined by the decomposition (\ref{Leffpert1}), again using dimensional
regularization to regularize both ultraviolet and infrared divergences.
This calculation is trivial, since massless loop diagrams with no external
legs vanish in dimensional regularization due to a cancellation between
ultraviolet poles in $\epsilon$ and infrared poles in $\epsilon$.
The result is therefore
\begin{equation}
{T \log {\cal Z} \over V} \;\approx\; - f \; T.
\label{logZeff2}
\end{equation}
Matching the results (\ref{logZ}) and (\ref{logZeff2}), we obtain
the coefficient $f$ to order $g^4$:
\begin{equation}
f \;=\; {\pi^2 \over 9} T^3
\Bigg\{ - {1 \over 10} \;+\; {1 \over 8} {g^2(\mu) \over 16 \pi^2}
\;-\; {1 \over 8}
\left[ 3 \log{\mu \over 4 \pi T} + {31 \over 15} + \gamma
+ 4 {\zeta'(-1) \over \zeta(-1)} - 2 {\zeta'(-3) \over \zeta(-3)} \right]
\left( {g^2 \over 16 \pi^2} \right)^2 \Bigg\} ,
\label{f}
\end{equation}
where $g^2(\mu)$ is the coupling constant in the $\overline{\rm MS}$
renormalization scheme at the scale $\mu$.
We have used the renormalization group equation (\ref{dg}) to replace
$\Lambda$ in (\ref{logZ}) by an arbitrary scale $\mu$ associated
with the renormalization of the full 4-dimensional theory.
Thus, at this order in $g^2$, the coefficient $f$ does not depend on
the ultraviolet cutoff $\Lambda$ of the effective theory.

\subsection{Mass parameter}

In this subsection, we calculate the coefficient $m^2(\Lambda)$
of the $\phi^2/2$ term in the effective lagrangian (\ref{Leff})
to next-to-leading order in $g^2$.  The parameter $m(\Lambda)$
can be interpreted as the contribution to the screening mass from
short distances of order $1/T$.  The actual screening mass $m_s$
is defined by the condition that the propagator for spacelike momentum
$K = (k_0=0,{\bf k})$ has a pole at ${\bf k}^2 = - m_s^2$.
The physical quantity $m_s^2$ coincides with $m^2(\Lambda)$
at order $g^2$, but $m_s^2$ has corrections of order $g^3$ which arise
from the long-distance scale $1/(gT)$.  In contrast, the mass parameter
$m^2(\Lambda)$ receives contributions only from the short-distance
scale $1/T$, and thus has a perturbative expansion in powers of $g^2(T)$.

One way to determine $m^2(\Lambda)$ is to match the propagator for the
zero-frequency mode of the field $\Phi({\bf x},\tau)$ in the full theory
with the propagator for $\phi({\bf x})$ in the effective theory.
At leading order in $g^2$, these operators are related as in (\ref{phi}).
This identification is sufficient for determining $m^2(\Lambda)$
to next-to-leading order in $g^2$.  Beyond that order, we must allow
for a short-distance coefficient multiplying $\phi({\bf x})$ in
(\ref{phi}), and we must also allow for the fact that $\phi({\bf x})$
is only the first term in an operator expansion that contains
$\phi^3({\bf x})$ and other higher dimension operators.  In general,
we must include all operators that are odd under $\phi \to - \phi$,
each multiplied by a short-distance coefficient.  In order to determine
$m^2(\Lambda)$ and all the necessary short-distance coefficients,
we would have to match not only the propagator but other 2-point
functions as well.

A simpler way to determine $m^2(\Lambda)$ is to match the screening
mass in the full theory and in the effective theory.  The screening mass
gives the location of the pole in the propagator for the
zero-frequency mode $\int_0^\beta d \tau \Phi({\bf x},\tau)$.
Denoting the self-energy function for the field $\Phi({\bf x},\tau)$ at
momentum $K=(k_0,{\bf k})$ by $\Pi(k_0,{\bf k})$, the screening
mass $m_s$ is the solution to the equation
\begin{equation}
k^2 + \Pi(0,{\bf k})\;=\;0\qquad \mbox{at $k^2=-m_s^2$}.
\label{msdef}
\end{equation}
The location of the pole is
independent of field redefinitions.  Since the operator expansion that
generalizes (\ref{phi}) can be interpreted as a field redefinition,
the screening mass $m_s$ also gives the location of the pole in the
propagator for the field $\phi({\bf x})$.  Denoting the self-energy for
$\phi({\bf x})$ by $\Pi_{\rm eff}(k,\Lambda)$, the screening mass $m_s$
must satisfy
\begin{equation}
k^2 + m^2(\Lambda) + \Pi_{\rm eff}(k,\Lambda)\;=\;0 \qquad
    \mbox{at $k^2=-m_s^2$}.
\label{msdefeff}
\end{equation}
By matching the expressions for $m_s$
obtained by solving (\ref{msdef}) and (\ref{msdefeff}),
we can determine the short-distance parameter $m^2(\Lambda)$.

We will obtain a perturbative expression for the screening mass $m_s$
in the full theory by calculating $\Pi(K)$ to order $g^4$ using the
strict perturbation expansion defined by the decomposition (\ref{Leffpert1}).
The self-energy is given by the sum of the Feynman diagrams in
Fig.~3:
\begin{eqnarray}
\Pi(K) \;\approx\; {Z^2_g g^2 \over 2}\;\sumint_P {1 \over P^2}
\;-\; {g^4 \over 4}\; \sumint_P {1 \over P^2} \;
	\sumint_P {1 \over (P^2)^2}
\;-\; {g^4 \over 6}\; \sumint_{PQ} {1 \over P^2 Q^2 (P+Q+K)^2} .
\label{Pi}
\end{eqnarray}
We can simplify the equation (\ref{msdef}) by expanding $\Pi(0,{\bf k})$
as a Taylor expansion around ${\bf k} = 0$.  This can be justified by the
fact that the leading order solution to (\ref{msdef}) gives a value
of $k$ that is of order $gT$, and $\Pi(K)$ is independent of $K$ at
leading order.  After setting $K=0$ in the last integral in
(\ref{Pi}), the solution to (\ref{msdef}) to order $g^4$ is trivial:
\begin{eqnarray}
m_s^2 \;\approx\;  {Z^2_g g^2 \over 2}\;\sumint_P {1 \over P^2}
\;-\; {g^4 \over 4}\; \sumint_P {1 \over P^2} \;
	\sumint_P {1 \over (P^2)^2}
\;-\; {g^4 \over 6}\; \sumint_{PQ} {1 \over P^2 Q^2 (P+Q)^2} .
\label{mspert}
\end{eqnarray}
It should be emphasized that this perturbative expression does not give a
physical value for the screening mass, because the sum-integrals are
infrared divergent.  However, as long as we can compute $m_s$
in the effective theory in a way that makes the same incorrect
assumptions about the long distance behavior, we can match the
perturbative expressions to determine the short-distance parameter
$m^2(\Lambda)$.

In order to match with the expression (\ref{mspert}), we have to
calculate the screening mass in the effective theory using the strict
expansion in $g^2$ defined by the decomposition (\ref{Leffpert1}).
The self-energy function $\Pi_{\rm eff}(k,\Lambda)$ in the equation
(\ref{msdefeff}) for the screening mass has a diagrammatic
expansion including the diagrams in Fig.~3 and the mass-insertion
diagrams in Fig.~4.  In the full theory, when $\Pi(K)$ is expanded
as a Taylor expansion around $K = 0$, terms that in dimensional
regularization scale like fractional powers of $k^2$ are automatically
set to 0.  We should therefore make the same simplifications in the
effective theory.  But then all the loop diagrams in Fig.~3
and Fig.~4 vanish,
since the external momentum ${\bf k}$ provides the only mass scale
in the integrals.  The self-energy function reduces to
\begin{equation}
\Pi_{\rm eff}(k,\Lambda) \;\approx\; \delta m^2,
\end{equation}
where $\delta m^2$ is the mass counterterm that contains the poles
in $\epsilon$ that are associated with mass renormalization.
The solution to the equation (\ref{msdefeff})
for the screening mass is therefore trivial:
\begin{equation}
m_s^2 \;\approx\; m^2(\Lambda) \;+\; \delta m^2.
\label{msperteff}
\end{equation}
{}From (\ref{msperteff}), we see that the
screening mass in this unphysical perturbation expansion is just the
bare mass. Comparing (\ref{mspert}) and
(\ref{msperteff}), we find that $m^2(\Lambda)$ is given by
\begin{equation}
m^2(\Lambda) \;=\;
{Z^2_g g^2 \over 2}\; \sumint_P {1 \over P^2}
\;-\; {g^4 \over 4}\; \sumint_P {1 \over P^2} \sumint_P {1 \over (P^2)^2}
\;-\; {g^4 \over 6}\; \sumint_{PQ} {1 \over P^2 Q^2 (P+Q)^2}
\;-\; \delta m^2,
\label{m2int}
\end{equation}
where the sum-integrals are to be evaluated using dimensional
regularization of both ultraviolet and infrared divergences.
The sum-integrals in (\ref{m2int}) are given in Appendix~A.
After renormalization of the
coupling constant $g$, there remains a pole in $\epsilon$ which must
be cancelled by the mass counterterm $\delta m^2$.  The mass
counterterm is thereby determined to be
\begin{equation}
\delta m^2 \;=\; {g^4 T^2 \over 24 (16 \pi^2)} {1 \over \epsilon}.
\label{deltam}
\end{equation}
Our final expression for the short-distance mass parameter
$m^2(\Lambda)$ is
\begin{equation}
m^2(\Lambda) = {1 \over 24} g^2(\mu) T^2 \;
\Bigg\{ 1
\;+\;  \left[ - 3 \log {\mu \over 4 \pi T}
	+ 4 \log {\Lambda \over 4 \pi T} + 2 - \gamma
	+ 2 {\zeta'(-1) \over \zeta(-1)} \right]
{g^2 \over 16 \pi^2} \Bigg\} ,
\label{m2}
\end{equation}
where $g^2(\mu)$ is the $\overline{\rm MS}$ coupling constant.
We have used the renormalization group equation (\ref{dg}) to
change the renormalization scale of the full theory from $\Lambda$
to $\mu$.  The remaining logarithm of $\Lambda$ in (\ref{m2})
reveals that $m^2(\Lambda)$ depends explicitly on the
factorization scale $\Lambda$ at order $g^4$.
This $\Lambda$-dependence is necessary to
cancel logarithmic ultraviolet divergences from loop integrals
in the effective theory.

\subsection{Coupling constants}

For the calculations in this paper, we require the coupling constant
$\lambda$ of the $\phi^4$ interaction in the effective theory only to
leading order in $g^2$.  At this order, we can simply read $\lambda$ off
from the lagrangian of the full theory.
Substituting $\Phi(\tau,{\bf x}) \longrightarrow \sqrt{T} \phi({\bf x})$ in
(\ref{L}) and comparing $\int_0^\beta d \tau {\cal L}$ with
${\cal L}_{\rm eff}$ in (\ref{Leff}), we find that, to leading order
in $g^2$,
\begin{equation}
\lambda \;=\; g^2 T .
\label{lambda}
\end{equation}
There is no dependence on the factorization scale $\Lambda$ at this order.
The coupling constant $\lambda$ could be calculated to higher order
in $g^2$ by matching 4-point correlation functions in the full theory
and in the effective theory.  Beyond next-to-leading order in $g^2$,
the matching is complicated by the breakdown of the simple
relation (\ref{phi}) between $\phi({\bf x})$ and the fundamental field.
A more convenient quantity for matching
beyond leading order is the on-shell scattering
amplitude defined by the residue of the 4-point function at the poles
of the propagators of the four external lines.  Like the screening mass,
this scattering amplitude is invariant under field redefinitions.

The only other coefficient in the effective lagrangian that is known is
the coefficient of $\phi^6$.  It has been computed
by Landsman \cite{landsman}, and its value is $15 \zeta(3) g^6/(128 \pi^4)$.
It first contributes to the free energy density at order $g^9$
and to the square of the screening mass at order $g^8$.

\begin{center} \section{Calculations in the Effective Theory} \end{center}

In this section, we calculate physical quantities in the effective theory
using perturbation theory.  We calculate the free energy to order $g^5$,
reproducing a recent result by Parwani and Singh \cite{parwani-singh},
and we obtain a new result for the screening mass to order $g^4$.

\subsection{Free Energy to Order $g^5$}

The free energy density $F(T)$ is defined by
\begin{equation}
{\cal Z}(T) \;=\; e^{- \beta \, F(T) \, V} .
\end{equation}
Comparing with the equivalent expression for the partition function
(\ref{Zeff}), we obtain
\begin{equation} \label{free}
F(T) = T \; f(\Lambda) - T \; {\log {\cal Z}_{\rm eff} \over V}\,,
\end{equation}
where the partition function for the effective theory is
\begin{equation}
{\cal Z}_{\rm eff} \;=\; \int^{(\Lambda)} {\cal D} \phi
\;\exp \left( - \int d^3x\; {\cal L}_{\rm eff} \right) .
\end{equation}
The strict perturbation expansion for $\log {\cal Z}_{\rm eff}$
corresponding to the decomposition (\ref{Leffpert1}) of ${\cal L}_{\rm eff}$
contains infrared divergences.  These divergences were not a problem
in the matching calculations of Section III,
since identical infrared divergences appeared
in the strict perturbation expansion for the full theory.
However, if we wish to actually calculate the free energy,
we must incorporate the physical effects that cut
off the infrared divergences into the free part of the lagrangian.
The necessary infrared cutoff is provided
by the $\phi^2$ term in the effective lagrangian.  We therefore make the
following decomposition of ${\cal L}_{\rm eff}$ into free and
interacting parts:
\begin{eqnarray}
\left({\cal L}_{\rm eff} \right)_{\rm free}
&=& {1 \over 2} \left( \makebox{\boldmath $\nabla$} \phi \right)^2
\;+\; {1 \over 2} m^2(\Lambda) \; \phi^2 ,
\nonumber \\
\left({\cal L}_{\rm eff} \right)_{\rm int}
&=&
{1 \over 4!} \lambda(\Lambda) \; \phi^4
\;+\; \delta {\cal L}.
\label{Leffpert2}
\end{eqnarray}
{}From the matching calculations in section III, $m^2$ is of order
$g^2T^2$ and $\lambda$ is of order $g^2T$.
Since the only momentum scale in $({\cal L}_{\rm eff})_{\rm free}$
is $m$, any powers of $T$ in the coefficient of an operator
will be compensated by powers of $m$.  Thus the effective expansion
parameter for the $\phi^4$ perturbation is $\lambda/m$, which is
of order $g$.  The next
most important perturbation after $\phi^4$ is the dimension-4 operator
$(\phi \makebox{\boldmath $\nabla$} \phi)^2$,
for which the effective expansion parameter
is of order $g^4 m/T$ or, equivalently, of order $g^5$.
Similarly, the effective expansion parameter for the dimension-3 operator
$\phi^6$ is $g^6$.  Thus to calculate
$\log {\cal Z}_{\rm eff}$ to next-to-next-to-leading order in $g$,
we need only consider the $\phi^4$ perturbations.

The contributions to $\log {\cal Z}_{\rm eff}$ of orders
$g^3$, $g^4$, and $g^5$
are given by the sum of the 1-loop, 2-loop, and 3-loop diagrams in
Fig.~1 and the first diagram in Fig.~2:
\begin{eqnarray}
{\log {\cal Z}_{\rm eff} \over V}
&=& - {1 \over 2} \int_p \log(p^2 + m_0^2)
\;-\; {\lambda \over 8} \left( \int_p {1 \over p^2 + m^2} \right)^2
\nonumber \\
&& \;+\; {\lambda^2 \over 16} \left( \int_p {1 \over p^2 + m^2} \right)^2
	\int_p {1 \over (p^2 + m^2)^2}
\nonumber \\
&& \;+\; {\lambda^2 \over 48}
\int_{pqr} {1 \over (p^2 + m^2) (q^2 + m^2) (r^2 + m^2)
[({\bf p}+{\bf q}+{\bf r})^2 + m^2]} \,.
\label{logZeffint}
\end{eqnarray}
The only effect of the diagram with the mass counterterm in
Fig.~2 is to replace $m^2$ in the 1-loop diagram
by $m_0 = m^2 + \delta m^2$, where the mass counterterm is
\begin{equation}
\delta m^2 \;=\; {2 \over 3} \left( \lambda \over 16 \pi \right)^2
{1 \over \epsilon}\,.
\label{barem}
\end{equation}
With the identification $\lambda = g^2 T$, this expression is
identical to the expression in (\ref{deltam}).
When $\int \log(p^2 + m_0^2)$ is expanded in powers of
$\lambda$ using~(\ref{barem}), the $1/\epsilon$ term cancels
against a pole from the last integral
in (\ref{logZeffint}), but it also gives rise to finite contributions
from the expansion of the integral $\int \log(p^2 + m_0^2)$ to order
$\epsilon$. The integrals in~(\ref{logZeffint}) are given in
Appendix~B. Adding up the diagrams, we obtain
\begin{equation}
{\log {\cal Z}_{\rm eff} \over V}
\;=\; {1 \over 12 \pi} m^3(\Lambda)
-\; {1\over 8\pi} {\lambda \over 16 \pi}m^2
\;-\; {1\over 12\pi} \left[
4 \log {\Lambda \over 2m} + {9\over 2} - 4 \log 2 \right]
	\left( {\lambda \over 16 \pi} \right)^2 m \,.
\label{logZeff3}
\end{equation}

Our final result for the free energy (\ref{free}) to
order $g^5$ is the sum of two terms that represent the contributions
from the momentum scales $T$ and $gT$, respectively:
\begin{equation}\label{free12}
  F(T)=F_1(T)+F_2(T) \,.
\end{equation}
The first term $F_1(T)=f T$ is the contribution
to the free energy from single-particle effects involving
the short-distance scale $1/T$ only. This contribution can
be expressed as a power series in $g^2(2\pi T)$, with the
first three terms given by~(\ref{f}):
\begin{eqnarray}
F_1(T) &=& - {\pi^2 \over 9} T^4\;
\Bigg\{
  {1 \over 10} \;-\; {1 \over 8} {g^2(2 \pi T) \over 16 \pi^2}
\nonumber\\
&& \qquad \qquad\;+\; {1 \over 8}
  \left[ {31 \over 15} - 3 \log 2 + \gamma +
  4 {\zeta'(-1) \over \zeta(-1)} - 2 {\zeta'(-3) \over \zeta(-3)} \right]
\left( {g^2 \over 16 \pi^2} \right)^2 \Bigg\}\,.
\label{F1}
\end{eqnarray}
The choice $\mu = 2 \pi T$ for the renormalization scale of the
effective theory will be justified in Section V.
The second term $F_2(T) = -T \log{\cal Z}_{\rm eff}/V$
in the free energy~(\ref{free12})
takes into account collective effects of the particles
involving the long-distance scale $1/(gT)$. It can be expressed
as a perturbation series in $\lambda$ and in the other
coupling constants of the effective theory. The first three
terms are given by~(\ref{logZeff3}):
\begin{equation}
F_2(T) \;=\;  \;-\; {1 \over 12 \pi} m^3(\Lambda) \; T\;
\Bigg\{ 1 \;-\; {3 \over 2} {\lambda \over 16 \pi m}
  \;-\; \left[4 \log {\Lambda \over 2 m} + {9\over 2} - 4\log 2\right]
  \left({\lambda\over 16\pi m}\right)^2 \Bigg\}\,,
\label{free2}
\end{equation}
where $\lambda=g^2 T$ and the mass parameter $m(\Lambda)$
is given by~(\ref{m2}) with $\mu = 2 \pi T$.
Note that, after substituting the expression (\ref{m2}) for the mass
parameter $m^2(\Lambda)$, the $\Lambda$-dependence in (\ref{free2})
cancels to next-to-next-to-leading order in $g$.
Adding the short-distance contribution in (\ref{F1}) and expanding
in powers of $g(2 \pi T)$, the free energy reduces to
\begin{eqnarray}
F(T) &=& - {\pi^2 \over 9} T^4\;
\Bigg\{
{1 \over 10} \;-\; {1 \over 8} \left( {g(2 \pi T) \over 4 \pi} \right)^2
\;+\; {1 \over \sqrt{6}} \left({g(2\pi T) \over 4\pi}\right)^3\,
\nonumber \\
&& \;+\; {1 \over 8}
  \left[ - {59 \over 15} - 3 \log 2 + \gamma +
  4 {\zeta'(-1) \over \zeta(-1)} - 2 {\zeta'(-3) \over \zeta(-3)} \right]
\left( {g \over 4 \pi} \right)^4 \Bigg\}
\nonumber\\
&& \;+\; \sqrt{3 \over 8}
\left[4 \log{g\over 4 \pi\sqrt{6}} - {5 \over 2} + 7 \log 2 - \gamma
    +2{\zeta'(-1)\over\zeta(-1)}\right]
	\left({g \over 4 \pi}\right)^5 \Bigg\}\,.
\label{freeg}
\end{eqnarray}
The coefficient of $g^4$ was first
calculated by Frenkel, Saa, and Taylor \cite{fst}, up to an error that
was corrected by Arnold and Zhai \cite{arnold-zhai}.  The order $g^5$
term was recently calculated by Parwani and
Singh \cite{parwani-singh}.  Our result agrees with theirs
after taking into account the difference in the definition of the
coupling constant.  There is a loss of accuracy when we make
a strict expansion in powers of $g$ as in (\ref{freeg}).
In Section V, we will give a more accurate expression for the free energy
which sums up leading logarithms of $g$ from higher orders
of perturbation theory.

\subsection{Screening Mass to Order $g^4$}

The screening mass $m_s$ describes the long-distance behavior
of the potential produced by the exchange of a particle with spacelike
momentum. The potential falls exponentially like $e^{-m_s R}$ at
large R. The screening mass is a long distance quantity, so it should
be calculable using the effective field theory. In this section, we
use the effective field theory to calculate $m_s$ to order $g^4$.

The screening mass $m_s$, which gives the location of the pole in the
propagator of the effective theory, is the solution to the equation
(\ref{msdefeff}).  At leading order in $\lambda/m$,
the solution is simply $m_s = m(\Lambda)$.
The self energy function $\Pi_{\rm eff}(k,\Lambda)$
is given to next-to-leading order in $\lambda$ by the Feynman
diagrams in Figs.~3 and~4:
\begin{eqnarray}
  \Pi_{\rm eff}(k,\Lambda) &=&
  {\lambda \over 2} \int_p {1 \over p^2 + m^2}\;
  -\; {\lambda^2 \over 4} \int_p {1 \over p^2 + m^2}
  \int_p {1 \over (p^2 + m^2)^2}
\nonumber \\
  && \qquad -\; {\lambda^2 \over 6}
  \int_{pq} {1 \over (p^2 + m^2) (q^2 + m^2)
  [({\bf p}+{\bf q}+{\bf k})^2 + m^2]}
\;+\; \delta m^2 \,.
\label{Pieff}
\end{eqnarray}
The mass counterterm $\delta m^2$, which is given in (\ref{barem}),
cancels an ultraviolet
pole in $\epsilon$ in the integral over $p$ and $q$. This is the only
integral in~(\ref{Pieff}) that depends on $k$. The self-consistent
solution to~(\ref{msdefeff}) to next-to-leading order in $\lambda/m$
is obtained by evaluating the integral at the point $k=im$. The
resulting expression for the screening mass is
\begin{eqnarray}
m_s^2 &=& m^2(\Lambda)
\;+\; {\lambda \over 2} \int_p {1 \over p^2 + m^2}
\;-\; {\lambda^2 \over 4} \int_p {1 \over p^2 + m^2}
	\int_p {1 \over (p^2 + m^2)^2}
\nonumber \\
&& \left. \qquad \;-\; {\lambda^2 \over 6} \int_{pq}
	{1 \over (p^2 + m^2) (q^2 + m^2)
	[({\bf p}+{\bf q}+{\bf k})^2 + m^2]}\right|_{k = im}
\;+\; \delta m^2 \,.
\label{scrm}
\end{eqnarray}
Note that the screening mass is not identical to the value of the inverse
propagator at 0 momentum, which would be given by (\ref{scrm}) with the last
integral evaluated at $k=0$.  Unlike the screening mass, the mass defined
by the inverse propagator at $k=0$ is not invariant under field redefinitions.

The integrals in (\ref{scrm}) are given in Appendix~B.
To next-to-next-to-leading order in $\lambda/m$, the screening mass is
\begin{equation}
  m_s^2\;=\;m^2(\Lambda) \Bigg\{ 1 \;-\; 2 \; {\lambda\over 16 \pi m}
\;-\; {2 \over 3} \left[4 \log{\Lambda\over 2m} + 3
  - 8 \log 2 \right]\left({\lambda\over 16 \pi m}\right)^2 \Bigg\} .
\label{ms2}
\end{equation}
Note that, when we substitute (\ref{m2}) for $m^2(\Lambda)$, the
$\Lambda$-dependence cancels to next-to-next-to-leading order in $g$.
This verifies that the screening mass is independent of the
arbitrary factorization scale $\Lambda$ to this order.
Setting $\mu = 2 \pi T$ in (\ref{m2}), substituting it into (\ref{ms2}),
and expanding in powers of $g(2 \pi T)$, we obtain
\begin{equation}
m_s^2 \;=\; {1\over 24} \; g^2(2 \pi T) \; T^2 \;
\Bigg\{1 - \sqrt{6}\; {g\over 4\pi}
\;+\; \left[4\log{g\over 4\pi\sqrt{6}}
    -1+ 11 \log 2-\gamma+2{\zeta'(-1)\over\zeta(-1)}\right]
    \left({g\over 4\pi}\right)^2\Bigg\}\,.
\label{ms}
\end{equation}
The term of order $g^3$ in (\ref{ms}) was first calculated by
Dolan and Jackiw \cite{dolan-jackiw}.  The correction of order $g^4$
in the expression for the screening mass is a new result.
The screening mass differs at order $g^4$ from the quasiparticle mass,
which was calculated to order $g^4$ by Parwani \cite{parwani1}.
The quasiparticle mass is defined by the pole in the energy for the
propagator at zero 3-momentum.  It gives the energy of the single-particle
excitations in the plasma, while the screening mass gives the range of
the force mediated by the exchange of a particle in the plasma.
There is a loss of accuracy in making a strict expansion of the
screening mass in powers of $g$ as in (\ref{ms}).  In Section V, we will
give a more accurate expression for the screening mass which resums
leading logarithms of $g$ from higher orders of perturbation theory.

\begin{center} \section{Summation of Leading Logarithms} \end{center}

The coefficients in the effective lagrangian (\ref{Leff}) depend on
not only on the coupling constant $g^2$ and the temperature $T$,
but also on two arbitrary momentum scales:  the renormalization scale
$\mu$ of the full theory and the ultraviolet cutoff $\Lambda$
of the effective theory.  In this section, we exploit these
arbitrary scales to sum up the leading logarithms from higher orders
in perturbation theory.  We first discuss the choice of the
renormalization scale $\mu$.  We then present the evolution equations that
describe the dependence of the short-distance coefficients on the scale
$\Lambda$.  We then show how the solutions of the evolution equations
can be used to sum up leading logarithms of $T/(gT)$
in physical quantities.

\subsection{Renormalization scale}

The short-distance coefficients in the effective lagrangian
would be independent of the arbitrary renormalization scale $\mu$ of the
original theory if they were calculated to all orders in $g^2$.
A dependence on $\mu$ appears however when a coefficient is calculated
only to a finite order in $g^2$.  At leading nontrivial order in $g^2$,
the scale $\mu$ appears to be completely arbitrary.  The resulting
ambiguity can be decreased by a next-to-leading order calculation,
because an inappropriate choice of $\mu$ will result in unnecessarily large
perturbative corrections.  The short-distance coefficients
$f$ and $m^2(\Lambda)$ were both calculated to order $g^4$ in Section III,
and we can use those results to discuss the appropriate scale for $\mu$
in the $g^2$ terms.

The transcendental constants appearing in the expressions (\ref{f})
and (\ref{m2}) have the numerical values
\begin{equation}
\gamma \;=\; 0.57722,
\qquad {\zeta'(-1) \over \zeta(-1)} \;=\; 1.98505,
\qquad {\zeta'(-3) \over \zeta(-3)} \;=\; 0.64543.
\end{equation}
The short-distance coefficients then reduce to
\begin{eqnarray}
f &=& - {\pi^2 \over 90} T^3
\Bigg\{ 1 \;-\; {5 \over 4} {g^2(\mu) \over 16 \pi^2}
\left[ 1 \;+\;
\left( - 7.2 - 3 \log{\mu \over 2 \pi T} \right)
	{g^2 \over 16 \pi^2} \right] \Bigg\} ,
\label{fnum} \\
m^2(\Lambda) &=& {1 \over 24} g^2(\mu) T^2 \;
\left[ 1 \;+\;
	\left( 4.7 - 3 \log {\mu \over 2 \pi T}
	+ 4 \log {\Lambda \over 2 \pi T} \right)
{g^2 \over 16 \pi^2} \right] ,
\label{m2num}
\end{eqnarray}
where the coupling constant is the $\overline{\rm MS}$ constant
at the scale $\mu$.  This coupling constant is defined so that
it is the appropriate coupling constant for particles whose
Euclidean invariant mass-squared $k_0^2 + k^2$
is approximately $\mu^2$.  The parameters (\ref{fnum}) and (\ref{m2num})
are coefficients in an effective lagrangian obtained
by integrating out modes with nonzero Matsubara frequencies
$k_0 = 2 n \pi T$, $n=1,2,3,\ldots$.  Since these modes have
invariant masses that are equal to or
greater than $2 \pi T$, we expect $\mu = 2 \pi T$
to be an appropriate choice for the renormalization scale.
Choosing $\mu = \Lambda = 2 \pi T$, the coefficients of $g^2/(16\pi^2)$
in square brackets in (\ref{fnum}) and in (\ref{m2num}) are
$-7.2$ and 4.7, respectively, which are reasonably small.
The more naive choice $\mu = \Lambda = T$
gives even smaller coefficients $-1.7$ and 2.9.
We prefer the physically motivated choice, and we therefore set
the renormalization scale to $\mu= 2 \pi T$ throughout the remainder
of this paper.

\subsection{Factorization scale}

The matching calculations described in Section III give the short-distance
coefficients in the effective lagrangian ${\cal L}_{\rm eff}$
as perturbation series in $g^2(2 \pi T)$
with coefficients that are polynomials in $\log(2 \pi T/\Lambda)$.
To avoid unnecessarily large coefficients in these perturbative expansions,
we must choose $\Lambda$ of order $2 \pi T$.
Once the short-distance coefficients are known, physical quantities can be
calculated in the effective theory using a perturbation expansion in
$\lambda/m$ and in other dimensionless parameters obtained by dividing the
short-distance parameters by appropriate powers of $m$.  The coefficients
in the resulting perturbation expansions for physical quantities
contain logarithms of $\Lambda/m$.
To avoid unnecessarily large coefficients in these perturbation expansions,
it is necessary to carry out the perturbative
calculations in the effective theory using short-distance parameters
that are evaluated at a scale $\Lambda$ of order $m$.
The parameters calculated at the original scale $\Lambda = 2 \pi T$
must therefore be evolved down to the scale $\Lambda = m$ before they
can be used in these perturbative calculations.
The $\Lambda$-dependence of these parameters is described by
``renormalization group equations'' or ``evolution equations''.

The effective lagrangian (\ref{Leff}) can be expressed as a sum over
over all local operators that respect the symmetries of the theory:
\begin{equation}
f(\Lambda) \;+\; {\cal L}_{\rm eff}
\;=\; \sum_n C_n(\Lambda) \; {\cal O}_n ,
\end{equation}
where we have included the unit operator as one of the operators ${\cal O}_n$.
The coefficients $C_n$ are the generalized coupling constants of the effective
theory. Because of ultraviolet divergences, the effective theory
must be regularized with an ultraviolet cutoff $\Lambda$.
The ultraviolet divergences in the effective theory include power
ultraviolet divergences proportional to $\Lambda^p$, $p=1,2,\ldots$,
and logarithmic divergences proportional to $\log(\Lambda/m)$.
As discussed in Section II, the power divergences are artifacts of the
regularization scheme and have no physical content.
If they are not removed as part of the regularization procedure,
they must be cancelled by power divergences in the coupling constants $C_n$.
In contrast, the logarithmic ultraviolet divergences
are directly related to logarithms of $T$ in the full theory,
and therefore represent real physical effects.
This difference justifies treating power ultraviolet divergences and the
logarithmic ultraviolet divergences differently.
It is convenient to use a regularization procedure for the effective theory
in which power ultraviolet divergences are subtracted and logarithmically
ultraviolet divergent integrals are cut off at the scale $\Lambda$.
These logarithmic divergences are then the only ones that must
be cancelled by the $\Lambda$-dependence of the coupling constants $C_n$.
The dimensions of a coupling constant can then only be taken up by powers
of the temperature $T$.  The coupling constant $C_n$ must be proportional
to $T^{3-d_n}$, where $d_n$ is the scaling dimension of the corresponding
operator ${\cal O}_n$.  The dimensionless factor multipling $T^{3-d_n}$
in the coupling constant $C_n$ can be computed
as a perturbation series in $g^2(T)$, with coefficients that are polynomials
in $\log(T/\Lambda)$. The dependence on $\Lambda$ is governed by a
``renormalization group equation'' or ``evolution equation'' of the form
\begin{equation}
\Lambda {d \ \over d \Lambda} C_n(\Lambda)
\;=\; \beta_n(C),
\label{dLeff}
\end{equation}
where the beta function $\beta_n$ has a power series expansion in the
coupling constants $C_m$.  These equations follow from the condition
that physical quantities must be independent of the arbitrary scale $\Lambda$.

Since $C_n$ is proportional to $T^{3-d_n}$,
every term in the expansion of its beta function must be proportional
to $T^{3-d_n}$. In particular, a term like $C_{m_1} C_{m_2} \ldots C_{m_k}$
can appear only if the dimensions $d_{m_i}$ of the corresponding operators
${\cal O}_{m_i}$ satisfy
\begin{equation}
\sum_{i=1}^k \left( 3 - d_{m_i} \right) \;=\; 3 - d_n.
\end{equation}
This condition is very restrictive, particularly if the effective field
theory is truncated to those terms that are given explicitly in (\ref{Leff}).
It implies that the beta function for the coefficient $f$ of the unit operator
can only have two terms proportional to $\lambda^3$ and $m^2 \lambda$.
The beta function for $m^2$ must be proportional to $\lambda^2$ and the
beta function for $\lambda$ must vanish to all orders in $\lambda$.
These restrictions reflect the super-renormalizability of this truncated
effective theory, which implies that there are only a finite number of
independent ultraviolet-divergent subdiagrams.

The evolution equations for $f$ and $m^2$ are calculated
in Appendix~C. They follow from the condition that the free energy and the
screening mass must be independent of $\Lambda$.  The evolution equations are
\begin{eqnarray}
\Lambda {d \ \over d \Lambda} f
&=& - {\pi \over 12} \left( {\lambda \over 16 \pi} \right)^3 ,
\label{df} \\
\Lambda {d \ \over d \Lambda} m^2
&=& {8 \over 3} \left( \lambda \over 16 \pi \right)^2,
\label{dm} \\
\Lambda {d \ \over d \Lambda} \lambda &=& 0.
\label{dlam}
\end{eqnarray}
The allowed term $m^2 \lambda$ does not appear in the beta function for $f$.
Using $\lambda = g^2 T$, we see that the beta function for $m^2$ in (\ref{dm})
is consistent with the explicit calculation to order $g^4$ in (\ref{m2}).
The evolution equations (\ref{df}), (\ref{dm}), and (\ref{dlam})
are correct to all orders in $\lambda$, but they receive corrections
involving the coefficients of higher dimension operators like $\phi^6$.
The coefficient of $\phi^6$ in the effective lagrangian is of order $g^6$.
It gives corrections to the right sides of the evolution equations
(\ref{df}), (\ref{dm}), and (\ref{dlam}) that are of order
$g^{12} \lambda^3$, $g^6 \lambda^2$, and $g^6 \lambda$, respectively.

To the accuracy given in (\ref{df}), (\ref{dm}), and (\ref{dlam}),
the solutions to the evolution equations are trivial.
The coupling constant $\lambda$ does not evolve with $\Lambda$, and the
solutions for $f$ and $m^2$ are simply linear in $\log(\Lambda)$:
\begin{eqnarray}
f(\Lambda) \;=\; f(2 \pi T) \;-\; {\pi \over 12}
	\left( {\lambda \over 16 \pi} \right)^3 \log(\Lambda/2 \pi T) ,
\label{dfsol}
\\
m^2(\Lambda) \;=\; m^2(2 \pi T) \;+\; {8 \over 3}
	\left( {\lambda \over 16 \pi} \right)^2 \log(\Lambda/2 \pi T).
\label{dmsol}
\end{eqnarray}
Note that, since the perturbative expression (\ref{m2}) is linear in
$\log(\Lambda)$, it already satisfies the evolution equation (\ref{dm}).

\subsection{Resumming Logarithms of $g$}

The evolution equations for the short-distance coefficients can be used
to sum up leading logarithms of $T/(gT)$ in physical quantities, such as
the free energy and the screening mass.  We must first choose a value
for the scale $\Lambda$ which will avoid unnecessarily large coefficients
in the perturbation expansions of the effective theory.  A reasonable
choice is the screening mass $m_s$, since this is the minimum invariant mass
for a particle in the 3-dimensional Euclidean theory.  At leading order in $g$,
the screening mass is simply $m_s = gT/\sqrt{24}$, so we choose
$\Lambda = g T/\sqrt{24}$.

An expansion for the screening mass in powers of $g$ is given in
(\ref{ms}).  This expression is only accurate up to corrections of order
$g^5 \log g$.  A more accurate expression for the screening mass
can be obtained by not expanding out the mass parameter in  (\ref{ms2}):
\begin{equation}
m_s^2 \;=\; m^2 \Bigg\{ 1 \;-\; 2 \; {g^2T \over 16 \pi m}
\;-\; {2 \over 3} \left[ 3
  - 12 \log 2 \right]\left({g^2T \over 16 \pi m}\right)^2 \Bigg\},
\label{msll}
\end{equation}
where $g = g(2 \pi T)$ is the coupling constant at the scale $2 \pi T$
and $m^2$ is the mass parameter in (\ref{m2}) with $\mu = 2 \pi T$ and
$\Lambda = gT/\sqrt{24}$:
\begin{equation}
m^2 \;=\; {1\over 24} g^2 \; T^2\; \Bigg\{ 1
\;+\; \left[ 4 \log {g\over 4\pi\sqrt{6}} + 2 - \log 2 -\gamma
  +2{\zeta'(-1)\over\zeta(-1)}\right] {g^2\over 16\pi^2}
\Bigg\}\,.
\label{m2m}
\end{equation}
The expression (\ref{msll}) is accurate up to corrections of order $g^5$.
It also sums up all the ``leading logarithms'' of order $g^{2n+3}\log^n g$,
$n = 2,3,\ldots$.  These terms are generated by expanding out the term
proportional to $g^2 T m$ in (\ref{msll}) using (\ref{m2m}).

An expansion for the free energy in powers of $g$ is given in
(\ref{freeg}).  This expression is accurate up to corrections of order
$g^6 \log g$.  A more accurate result can be obtained by using the solution
to the renormalization group equation (\ref{df}) for $f$ in the
short-distance part and by not expanding out
the mass parameter $m$ in the long-distance part (\ref{free2}):
\begin{eqnarray}
F(T) &=&  F_1(T)
\;-\; {\pi \over 12} \left( g^2 \over 16 \pi \right)^3 T^4
	\log {g \over 4 \pi \sqrt{6}}
\nonumber \\
&& \;-\; {1 \over 12 \pi} \; m^3 \; T\;
\Bigg\{ 1 \;-\; {3 \over 2} {g^2 T \over 16 \pi m}
  \;-\; \left[{9\over 2} - 8\log 2\right]
  \left({g^2 T \over 16\pi m}\right)^2 \Bigg\}\,,
\label{freell}
\end{eqnarray}
where $g = g(2 \pi T)$, and $m$ is given by (\ref{m2m}).
The first two terms on the right side of (\ref{freell}) are the
short distance contribution $f(\Lambda)T$, with the factorization scale
evaluated at $\Lambda = gT/\sqrt{24}$.  Its expansion to order $g^4$
is given by $F_1(T)$ in (\ref{F1}), and the term of order $g^6 \log g$
comes from the solution (\ref{dfsol}) to the renormalization group
equation for $f$.  There is also another contribution of order
$g^6 \log g$ that comes from expanding out the $g^2 m^2 T^2$ term
in (\ref{freell}) using (\ref{m2m}).
Having included both of these $g^6 \log g$ terms, the
expression (\ref{freell}) for the free energy is accurate
up to corrections of order $g^6$.  It also sums up all the
``leading logarithms'' of the form $g^{2n+3}\log^n g$, $n = 2,3,\ldots$,
which are obtained by expanding out the $m^3 T$ term in (\ref{freell}).

\begin{center} \section{Conclusions} \end{center}

We have developed an effective-field-theory approach for calculating the
thermodynamic properties of a field theory in the high temperature limit.
The effective field theory is the 3-dimensional theory obtained by
dimensional reduction to the bosonic zero-frequency modes.
The short-distance coefficients in the effective lagrangian are
computed by straightforward perturbative calculations in the full theory
without any resummation.  Thermodynamic quantities are then calculated
using a perturbation expansion in the effective theory which incorporates
the effects of screening.  In each of these two steps, the calculations
involve only a single mass scale, which greatly simplifies the sums and
integrals that need to be evaluated.  The short-distance coefficients
satisfy renormalization group equations which can be used to
improve the perturbation expansion by summing up leading logarithms
of $T/(gT)$.  The power of this method was demonstrated by carrying out
two calculations in massless $\Phi^4$ theory beyond the highest
orders that were previously available.  The free energy was calculated
to order $g^6 \log g$ and the screening mass was calculated to order
$g^5 \log g$.

The method that we have used to calculate the free energy of a scalar
field theory can also be applied to nonabelian gauge theories, such as QCD.
The free energy for QCD has been calculated with an error of order
$g^5 \log g$ using more conventional resummation methods \cite{arnold-zhai}.
Using our effective field theory approach, it should be straightforward
to decrease the error to order $g^6$.  The effective theory obtained by
integrating out the scale $T$ is a 3-dimensional gauge theory with a
scalar field in the adjoint representation.  There are two short distance
coefficients that must be calculated beyond leading order in $g^2$
in order to compute the free energy to order $g^5$.  The coefficient $f$
of the unit operator is required to next-to-next-to-leading order in $g^2$,
but this is already known \cite{braaten}.  The electric mass parameter
$m_{\rm el}^2$ must be calculated to next-to-leading order in $g^2$.
The error can be decreased further by calculating the renormalization group
equations for $f$ amd $m_{\rm el}^2$.  Their solutions can be used to sum
up terms of the form $g^6 \log^2 g$ and $g^6 \log g$,
thereby reducing the error to order $g^6$.
This is the maximal accuracy that can be achieved using purely diagrammatic
methods.  At order $g^6$, there is a contribution to the free energy
from the momentum scale $g^2T$ that can only be calculated using
lattice simulations of 3-dimensional QCD \cite{braaten}.

A similar effective field theory approach was recently developed by
Farakos, Kajantie, Rummukainen, and Shaposhnikov to study the electroweak
phase transition \cite{fkrs}.  They integrated out the scale $T$
to get a 3-dimensional effective theory, and exploited the
renormalization group equations of the effective theory to take into
account logarithms of $T/(gT)$, just as we have done in this paper.
They also integrating out the scale $gT$ to obtain a second effective field
theory that must be treated numerically.  This second step is necessary
in gauge theories since there is no screening of magnetostatic forces
at the scale $gT$.  This strategy has also been used to determine
the asymptotic behavior of the correlator of Polyakov loops
\cite{braaten-nieto} and to solve the problem of calculating the
magnetostatic contribution to the free energy of a
nonabelian gauge theory \cite{braaten}.

An effective field theory approach has also been applied recently to
the massless $\Phi^4$ theory by Marini and Burgess \cite{marini-burgess}.
They used a momentum cutoff as a regulator and their short-distance
coefficients therefore contain power ultraviolet divergences
that serve simply to cancel the power ultraviolet divergences from
loop integrals.  These power divergences greatly complicate the
renormalization group equations for the short-distance coefficients.
As we have emphasized, the power divergences are artifacts of the
regulator and might as well be subtracted as part of the regularization
scheme.  This not only greatly streamlines explicit calculations,
but it also makes the conceptual framework more transparent.

Effective field theories obtained by dimensional reduction have provided
great insight into the qualitative behavior of field theories in the
high temperature limit.   In this paper, we have shown how they can also
be used as a practical tool for explicit calculations.  By using effective
field theory to separate the important momentum scales $T$ and $gT$
(and $g^2T$ if necessary), perturbative calculations can be organized
into steps that involve only a single momentum scale at a time.
By exploiting the renormalization group structure of the effective theory
to sum up leading logarithms of $g$, potentially large coefficients
in the perturbative
expansion can be brought under control.  We have exhibited the power of
our effective-field-theory method by carrying out pioneering calculations
in massless $\Phi^4$ theory.   This method has many other exciting
applications, especially in unravelling the complexities of nonabelian
gauge theories at high temperature.

\bigskip

\section*{Acknowledgements}

This work was supported in part by the U.~S. Department of Energy,
Division of High Energy Physics, under Grant DE-FG02-91-ER40684,
and by the Ministerio de Educaci\'on y Ciencia of Spain.
One of us (E.B.) would like to thank P. Arnold and G. Bodwin for
valuable discussions.

\appendix\bigskip\renewcommand{\theequation}{\thesection.\arabic{equation}}
\begin{center}\section{Sum-integrals in the Full Theory}\end{center}
\setcounter{equation}{0}

In the imaginary-time formalism for thermal field theory, a boson has
Euclidean 4-momentum $P=(p_0,{\rm\bf p})$, with $P^2=p_0^2+{\rm\bf p}^2$. The
Euclidean energy $p_0$ has discrete values: $p_0=2\pi nT$, where $n$ is
an integer. Loop diagrams involve sums over $p_0$ and integrals over
{\bf p}. It is convenient to introduce a concise notation for these
sums and integrals.
If dimensional regularization is used to regularize ultraviolet or
infrared divergences, the definition is
\begin{equation}
  \sumint_P \;\equiv\;
  \left(\frac{e^\gamma\mu^2}{4\pi}\right)^\epsilon\;
  T\sum_{p_0}\:\int {d^{3-2\epsilon}p \over (2 \pi)^{3-2\epsilon}}\,,
\end{equation}
where $3-2\epsilon$ is the dimension of space and $\mu$ is an arbitrary
momentum scale. The factor $(e^\gamma/4\pi)^\epsilon$
is introduced so that, after minimal subtraction of the poles in $\epsilon$
due to ultraviolet divergences, $\mu$ coincides with the renormalization
scale of the $\overline{\rm MS}$ renormalization scheme.

The sum-integrals required to calculate the coefficient $f(\Lambda)$ to
next-to-leading order in $g^2$ can be found in
Ref.~\cite{arnold-zhai}. We reproduce them here for convenience:
\begin{eqnarray}
  \sumint_P\log P^2&=&-{\pi^2T^4 \over 45}
  \left[1+O(\epsilon)\right]\,,
\\
  \sumint_P\frac{1}{P^2}&=&\frac{T^2}{12}
  \left[1+\epsilon\left(
    2\log\frac{\mu}{4\pi T}+2+2\frac{\zeta'(-1)}{\zeta(-1)}
  \right)+O(\epsilon^2)\right]\,,
\\
  \sumint_P\frac{1}{(P^2)^2}&=&\frac{1}{(4\pi)^2}
  \left[\frac{1}{\epsilon}+2\log\frac{\mu}{4\pi T}+2\gamma+O(\epsilon)
  \right]\,,
\\
  \sumint_{PQ}{1\over P^2Q^2(P+Q)^2}& =& 0\,,
\\
  \sumint_{PQR}\frac{1}{P^2 Q^2 R^2 (P+Q+R)^2}&=&
    {T^4 \over 24(4\pi)^2}\left[
    \frac{1}{\epsilon}+6\log\frac{\mu}{4\pi T}\right.  \nonumber
\\
  && \qquad \qquad \left.+\frac{91}{15}+8\frac{\zeta'(-1)}{\zeta(-1)}-
      2\frac{\zeta'(-3)}{\zeta(-3)}+O(\epsilon)\right]\,,
\end{eqnarray}
where $\gamma$ is Euler's constant and $\zeta(z)$ is Riemann's
zeta function.

\begin{center}\section{Integrals in the Effective Theory}\end{center}
\setcounter{equation}{0}

The effective theory for the scale $g^2T$ is an Euclidean field theory
in 3 space dimensions. Loop diagrams involve integrals over 3-momenta.
It's convenient to introduce the notation $\int_p$ for these integrals.
If dimensional regularization in $3-2\epsilon$ dimensions is used to
regularize ultraviolet divergences, we use the integration measure
\begin{equation}
  \int_p\;\equiv\;
  \left(\frac{e^\gamma\mu^2}{4\pi}\right)^\epsilon\,
  \int {d^{3-2\epsilon}p \over (2 \pi)^{3-2\epsilon}}\,.
\end{equation}
If renormalization is accomplished by the minimal substraction of poles
in $\epsilon$, then $\mu$ is the renormalization scale in the
$\overline{\rm MS}$ scheme.

The integrals that are required to calculate the free energy to
order $g^5$ and the screening mass to order $g^4$ are
\begin{eqnarray}
  \int_p\log(p^2+m^2) & = & -{m^3\over 6\pi}
  \left[1 + \epsilon\left(2\log{\mu \over 2m} + {8 \over 3}\right)
  + O(\epsilon^2)\right]\,,
\label{bilog}
\\
  \int_p{1 \over p^2+m^2} & = & -{m\over 4\pi}
  \left[1 + \epsilon\left(2\log{\mu \over 2m} + 2\right)
  +O(\epsilon^2)\right]\,,
\label{bi1}
\\
\int_p{1 \over (p^2+m^2)^2} & = & {1\over 8\pi m}
	\left[1 + \epsilon\left(\,2\log{\mu \over 2m}\right)
	+ O(\epsilon^2)\right]\,,
\label{bi2}
\\
\lefteqn{
\int_{pq} {1\over p^2+m^2} {1\over q^2+m^2} {1\over ({\bf p}+{\bf q})^2+m^2}
} \hspace{1.5in}
\nonumber \\
& = & {1\over (8\pi)^2}
\left[ {1\over\epsilon} + 4\log{\mu\over 2m} + 2
	+ 4  \log 2 - 4 \log 3 + O(\epsilon) \right]\,,
\label{bi30}
\\
\lefteqn{
\left. \int_{pq} {1 \over p^2+m^2} {1\over q^2+m^2}
	{1 \over ({\bf p}+{\bf q}+{\bf k})^2 + m^2}\right|_{k=im}
}\hspace{1.5in}\nonumber
\nonumber \\
& = & {1\over (8\pi)^2}
\left[{1\over\epsilon} + 4\log{\mu\over 2m} + 6 - 8\log 2
	+ O(\epsilon)\right]\,,
\label{bi33}
\\
\lefteqn{
\int_{pqr}{1\over p^2+m^2}{1\over q^2+m^2}{1\over r^2+m^2}
	{1 \over ({\bf p}+{\bf q}+{\bf r})^2+m^2}
}\hspace{1.5in}\nonumber
\\
& = & -{m\over (4\pi)^3}
\left[{1\over\epsilon} + 6\log{\mu\over 2m} + 8 - 4\log 2
	+ O(\epsilon)\right] ,
\label{bi3}
\end{eqnarray}

The integral (\ref{bilog}) is standard.
The remaining integrals are most easily evaluated by going to coordinate
space. The Fourier transform of the propagator $1/(k^2+m^2)$ in
$3-2\epsilon$ dimensions defines a potential $V(R)$:
\begin{equation}
  V(R)\equiv\int_k e^{i{\rm\bf k}\cdot{\rm\bf R}}\;{1\over k^2+m^2}\,.
\end{equation}
It can be expressed in terms of the modified Bessel function $K_\nu(z)$:
\begin{equation}
  V(R)=\left({e^\gamma \mu^2 \over 4\pi}\right)^\epsilon
    {1\over (2\pi)^{3/2-\epsilon}}\left({m\over R}\right)^{1/2-\epsilon}
    K_{1/2-\epsilon}(mR)\,.
\end{equation}
At $\epsilon=0$, this reduces to the familiar Coulomb potential
\begin{equation}
  V_0(R)={e^{-mR}\over 4\pi R}\,.
\end{equation}
For small $R$, the potential $V(R)$ can be expressed as the sum of
two Laurent expansions in $R^2$, a singular one beginning with an
$R^{-1+2\epsilon}$ term and a regular one begining with an $R^0$ term:
\begin{eqnarray}
  V(R) & = &
    \left({e^\gamma\mu^2\over 4}\right)^\epsilon
    {\Gamma({1\over 2}-\epsilon)\over\Gamma({1\over 2})}
      {1\over 4\pi}R^{-1+2\epsilon}\left[1+{m^2R^2\over2(1+2\epsilon)}+
        O(m^4R^4)\right]   \nonumber
\\ \label{bex}
  &&
    -(e^\gamma\mu^2)^\epsilon
    {\Gamma(-{1\over 2}+\epsilon)\over\Gamma(-{1\over 2})}
      {1\over 4\pi}m^{1-2\epsilon}\left[1+{m^2R^2\over2(3-2\epsilon)}+
        O(m^4R^4)\right]\,.
\end{eqnarray}

The integrals~(\ref{bi1}) and~(\ref{bi2}) are related to the potential at
$R=0$.
By the rules of dimensional regularization, they can be evaluated by taking
$\epsilon$ large enough that the $R^{-1+2\epsilon}$ term vanishes as
$R\rightarrow 0$, and then analytically continuing back to small values of
$\epsilon$. The two integrals are
\begin{eqnarray}
  \int_p{1\over p^2+m^2} & \equiv & V(0)
    \;=\; -(e^\gamma \mu^2)^\epsilon
      {\Gamma(-{1\over 2}+\epsilon)\over\Gamma(-{1\over 2})}
      {1\over 4\pi}m^{1-2\epsilon}\,,
\\
  \int_p{1\over(p^2+m^2)^2} & \equiv &
  -{1\over 2m}\left.{d\over dm}V(R)\right|_{R=0}
    \;=\; (e^\gamma\mu^2)^\epsilon
      {\Gamma({1\over 2}+\epsilon)\over\Gamma({1\over 2})}
      {1\over 8\pi}m^{-1-2\epsilon}\,.
\label{B4exact}
\end{eqnarray}
The integrals~(\ref{bi33}) and~(\ref{bi3}) require a little
more effort. The integral~(\ref{bi3}) can be written
\begin{equation}\label{pot}
  \int_{pqr}{1\over p^2+m^2}{1\over q^2+m^2}{1\over r^2+m^2}
  {1\over ({\bf p}+{\bf q}+{\bf r})^2+m^2}=\int_R\,V^4(R)\,,
\end{equation}
where $\int_R$ is defined by
\begin{equation}
  \int_R\,\equiv\,\left({e^\gamma \mu^2 \over 4\pi}\right)^{-\epsilon}
  \int d^{3-2\epsilon}R\,\,.
\end{equation}
{}From the $R\rightarrow 0$ region of the integral~(\ref{pot}), there is a
linear
ultraviolet divergence, which is removed by dimensional regularization, and
a  logarithmic divergence, which appears as a pole in $\epsilon$. We evaluate
the integral by splitting the radial integration into two regions, $0<R<r$
and $r<R<\infty$. Since the ultraviolet divergences come only from the
region $R\rightarrow 0$, we can  set $\epsilon=0$ in the region $r<R<\infty$.
Thus the integral can be written
\begin{equation}
  \int_R\,V^4(R) \;=\; \left({e^\gamma \mu^2 \over 4}\right)^{-\epsilon}
    {\Gamma({3\over 2})\over\Gamma({3\over 2}-\epsilon)}
      \; 4\pi\int_0^r dR\,R^{2-2\epsilon}\,V^4(R)
	+\; 4\pi\int_r^\infty dR\,R^2\,V_0^4(R)\,.
\label{biv}
\end{equation}
By choosing $r\ll 1/m$, we can evaluate the first integral on the right side
of~(\ref{biv}) by using the small-$R$ expansion~(\ref{bex}) for $V(R)$.
Dropping all terms that vanish as $r\rightarrow 0$, we get
\begin{eqnarray}
  \lefteqn{
    \left({e^\gamma \mu^2 \over 4}\right)^{-\epsilon}
    {\Gamma({3\over 2})\over\Gamma({3\over 2}-\epsilon)}
    \; 4\pi\int_0^r dR\,R^{2-2\epsilon}\,V^4(R)
  }\hspace{1in}\nonumber
\\ \label{biv1}
  & = & {1\over (4\pi)^3}
    \left[-{1\over r} -
      m\left({1\over\epsilon}+2\log{r^2\mu^3\over 2m}+4\gamma+4\right)
    \right] + O(\epsilon)\,.
\end{eqnarray}
The  integral over the region $r<R<\infty$ is easily evaluated using
integration by parts. Dropping all terms that vanish as $r\rightarrow 0$,
we get
\begin{equation}\label{biv2}
  4\pi\int_r^\infty dR\,R^2\,V^4(R)={1\over(4\pi)^3}
    \left[{1\over r}+4m\left(\log 4mr+\gamma-1\right)\right]\,.
\end{equation}
Note that the $1/r$ and $\log r$ terms cancel between~(\ref{biv1})
and~(\ref{biv2}). Inserting~(\ref{biv1}) and~(\ref{biv2}) into~(\ref{biv}),
we obtain~(\ref{bi3}).

The integral~(\ref{bi33}) can be evaluated in a similar way
to~(\ref{bi3}). It can be written
\begin{equation}\label{pot3}
  \int_{pq}{1\over p^2+m^2}{1\over q^2+m^2}
  {1\over ({\bf p}+{\bf q}+{\bf k})^2+m^2}=
  \int_R\,e^{i{\bf k}\cdot{\bf R}}\;V^3(R)\,.
\end{equation}
Again we split the radial integration into two regions and set
$\epsilon=0$ in the region $r<R<\infty$. After evaluating the
angular integrals, we obtain
\begin{eqnarray}
  \int_R\,e^{i{\bf k}\cdot{\bf R}}\;V^3(R)&=&
    \left({e^\gamma \mu^2 \over 2k}\right)^{-\epsilon}
    {(2\pi)^{3/2}\over\sqrt{k}}\int_0^r dR\; R^{3/2-\epsilon}
    J_{1/2-\epsilon}(kR)\;V^3(R)
\nonumber\\ \label{intev3}
  &&\qquad \;+\; {4\pi\over k}\int_r^\infty dR\; R \sin(kR)\;V_0^3(R) \,,
\end{eqnarray}
where $J_\nu(z)$ is an ordinary Bessel function. We evaluate
the first integral using the small-$R$ expansion~(\ref{bex}) for
$V(R)$ and the small-$R$ expansion for the Bessel function
\begin{equation}
  J_{1/2-\epsilon}(kR)={1\over\Gamma({3\over 2}-\epsilon)}
    \left({1\over 2}kR\right)^{1/2-\epsilon}\left[1+O(k^2R^2)\right]\,.
\end{equation}
Dropping terms that vanish as $r\rightarrow 0$, the first integral
in~(\ref{intev3}) is
\begin{eqnarray}
 \left({e^\gamma \mu^2 \over 2k}\right)^{-\epsilon}
    {(2\pi)^{3/2}\over\sqrt{k}}\int_0^r dR\; R^{3/2-\epsilon}
    J_{1/2-\epsilon}(kR)\;V^3(R)
\nonumber \\
\qquad \qquad \qquad \qquad \qquad
\;=\; {1\over (8\pi)^2}\left({1\over\epsilon}+
    4\log\mu r + 2 + 4\gamma\right)\;+\;O(\epsilon) \,.
\label{intev31}
\end{eqnarray}
This integral is independent of $k$. In the second integral
in~(\ref{intev3}), we have to set $k=im$:
\begin{equation}
\left. {4\pi\over k}\int_r^\infty dR\; R \sin(kR)\;V_0^3(R) \right|_{k=im}
\;=\; {1\over (4\pi)^2}\left(-\log 2mr + 1 - 2\log 2 -
    \gamma\right)\,.
\label{intev32}
\end{equation}
Adding~(\ref{intev31}) and~(\ref{intev32}), the logarithms of $r$
cancel and we obtain~(\ref{bi33}).

\begin{center}\section{Evolution Equations for Coefficients}\end{center}
\setcounter{equation}{0}

In this appendix, we calculate the evolution equations for the
short-distance coefficients $f$ and $m^2$ in the effective lagrangian.
These equations follow from the condition that physical quantities
must be independent of the arbitrary renormalization scale $\Lambda$
of the effective theory.  The $\Lambda$-dependence of the parameters
in the effective lagrangian must cancel the $\Lambda$-dependence
from loop integrals.  If power ultraviolet divergences are subtracted as part
of the regularization scheme, then the $\Lambda$-dependence comes only from
logarithmically-divergent loop integrals.
One-loop diagrams in a 3-dimensional field theory never give logarithmic
ultraviolet divergences. After averaging over angles, such an integral has
the behavior $\int d^3p/(p^2)^n$ at large $p$. This gives a power divergence
for $n=1$ or less and is convergent for $n=2$ or greater. Thus logarithmic
divergences only arise in diagrams with 2 or more loops.  If the number
of loops is odd, logarithmic divergences only arise from subdiagrams
with an even number of loops.  Thus the evolution equations are completely
determined by diagrams with an even number of loops.

The renormalization group equations for $m^2$ can be determined from
the condition that the screening mass $m_s$ is independent of $\Lambda$.
Since only the ultraviolet region of loop integrals is relevant for
determining the evolution equations, we can use the expression for the
screening mass that is obtained from the strict perturbation expansion
in $g^2$ defined by the decomposition (\ref{Leffpert1}).  The expression
for the screening mass to order $g^4$ is
\begin{equation}
m_s^2 \;\approx\; m^2 \;+\; {\lambda \over 2} \int_p {1 \over p^2}
\;-\; {\lambda^2 \over 4} \int_p {1 \over p^2} \int_p {1 \over (p^2)^2}
\;-\; {\lambda^2 \over 6} \int_{pq} {1 \over p^2 q^2 ({\bf p} + {\bf q})^2}
\;-\; {\lambda m^2 \over 2} \int_p {1 \over (p^2)^2}
\;+\; \delta m^2 .
\end{equation}
The only logarithmic divergence comes from the 2-loop integral over $p$ and
$q$, which comes from the last diagram in Fig.~3.  Thus the condition that
$m_s^2$ is independent of $\Lambda$ reduces to
\begin{equation}
\Lambda {d \ \over d \Lambda} m^2 \;=\; {\lambda^2 \over 6} \;
\Lambda {d \ \over d \Lambda} \int_{pq} {1 \over p^2 q^2 ({\bf p} + {\bf
q})^2}.
\label{dmint}
\end{equation}
The ultraviolet divergence in the integral on the right side of
(\ref{dmint}) is the same as in (\ref{bi30}).
The derivative is with respect to the scale $\Lambda$ associated with
the ultraviolet cutoff, and must be taken with the infrared cutoff fixed.
The integral in (\ref{dmint}) vanishes in dimensional regularization
only if we use the same
regularization scale $\mu$ for ultraviolet and infrared divergences.
If we use a different scale $\Lambda$ for the regularization
of ultraviolet divergences, the integral is
\begin{equation}
\int_{pq} {1 \over p^2 q^2 ({\bf p}+{\bf q})^2}
\;=\; {1 \over (8\pi)^2}
\left[ \left({1\over\epsilon}\right)_{UV}
	- \left({1\over\epsilon}\right)_{IR}
	+ 4\log{\Lambda \over \mu} \right]\,.
\label{intpq}
\end{equation}
The subscripts $UV$ and $IR$ indicate whether the pole
in $\epsilon$ is of ultraviolet or infrared origin.  Inserting
(\ref{intpq}) into (\ref{dmint}), we obtain the evolution equation
\begin{equation}
\Lambda {d \ \over d \Lambda} m^2 \;=\; {8 \over 3}
\left( \lambda \over 16 \pi \right)^2 .
\label{m2evol}
\end{equation}
This result is accurate to all orders in $\lambda$ and to leading order
in the coefficients of higher dimension operators.

The evolution equation for $f$ can be determined from the condition that the
free energy, or equivalently, the logarithm of the partition function
given in (\ref{logZeff1}), is independent of $\Lambda$.
We need only consider diagrams for $\log {\cal Z}_{\rm eff}$
which have an even number of loops, and they can be calculated using the
strict perturbation expansion in $g^2$ defined by the decomposition
(\ref{Leffpert1}).  The two-loop diagram for $\log {\cal Z}_{\rm eff}$
in Fig.~1 has no logarithmic ultraviolet divergence.
The four-loop diagrams that have logarithmic ultraviolet
divergences are shown in Fig.~5. The first diagram in Fig.~5 has two-loop
subdiagrams that are logarithmically divergent. The resulting
$\Lambda$-dependence is cancelled by the $\Lambda$-dependence of
$m^2$ in the coefficient of the second diagram of Fig.~2.
The second diagram in Fig.~5
has no logarithmically divergent subdiagrams, but it has an overall
logarithmic divergence.  The $\Lambda$-dependence of this contribution
to the free energy can only be cancelled by that of the short-distance
coefficient $f$:
\begin{equation}
\Lambda {d \ \over d \Lambda} f
\;=\; - {\lambda^3 \over 48} \Lambda {d \ \over d \Lambda}
\int_{pq_1q_2q_3} {1 \over q_1^2 ({\bf p} + {\bf q}_1)^2
	q_2^2 ({\bf p} + {\bf q}_2)^2 q_3^2 ({\bf p} + {\bf q}_3)^2}.
\end{equation}
After combining pairs of propagators using the Feynman parameter trick,
the integrals over $q_1$, $q_2$, and $q_3$ can be evaluated analytically
using (\ref{B4exact}).  The result is
\begin{eqnarray}
&& \int_{pq_1q_2q_3} {1 \over q_1^2 ({\bf p} + {\bf q}_1)^2
	q_2^2 ({\bf p} + {\bf q}_2)^2 q_3^2 ({\bf p} + {\bf q}_3)^2}
\nonumber \\
&& \quad \quad \quad \quad
\;=\; \left[ {1 \over 8 \pi} e^{\gamma \epsilon} \mu^{2 \epsilon}
	{\Gamma({1 \over 2} + \epsilon) \over \Gamma({1 \over 2})}
	\int_0^1 dx (x-x^2)^{-{1\over 2} - \epsilon} \right]^3
	\int_p p^{-3 - 6 \epsilon}
\label{dfint}
\end{eqnarray}
The integral over $p$ vanishes in dimensional regularization if the same
scale $\mu$ is used in the regularization of ultraviolet and infrared
divergences.  If a different scale $\Lambda$ is used for ultraviolet
divergences, the value of the integral is
\begin{equation}
\int_{pq_1q_2q_3} {1 \over q_1^2 ({\bf p} + {\bf q}_1)^2
	q_2^2 ({\bf p} + {\bf q}_2)^2 q_3^2 ({\bf p} + {\bf q}_3)^2}
\;=\; {1 \over 32 (16 \pi)^2}
\left[ \left({1\over\epsilon}\right)_{UV}
	- \left({1\over\epsilon}\right)_{IR}
	+ 8 \log{\Lambda \over \mu} \right]\,,
\end{equation}
Inserting this result into (\ref{dfint}), we obtain the evolution equation
\begin{equation}
\Lambda {d \ \over d \Lambda} f \;=\; - {\pi \over 12}
\left( \lambda \over 16 \pi \right)^3 .
\label{fevol}
\end{equation}
This result is accurate to all orders in $\lambda$ and to leading order
in the coefficients of higher dimension operators.


\newpage
\begin{center}\section*{Figure Captions}\end{center}

\begin{enumerate}

\item
Feynman diagrams for the logarithm of the partition function
in the full theory and in the effective theory.

\item
Additional Feynman diagrams for the logarithm of the partition function
in the effective theory.

\item
Feynman diagrams for the self-energy
in the full theory and in the effective theory.

\item
Additional Feynman diagram for the self-energy in the effective theory.

\item
Four-loop diagrams for the logarithm of the partition function of
the effective theory which have logarithmic ultraviolet divergences.

\end{enumerate}

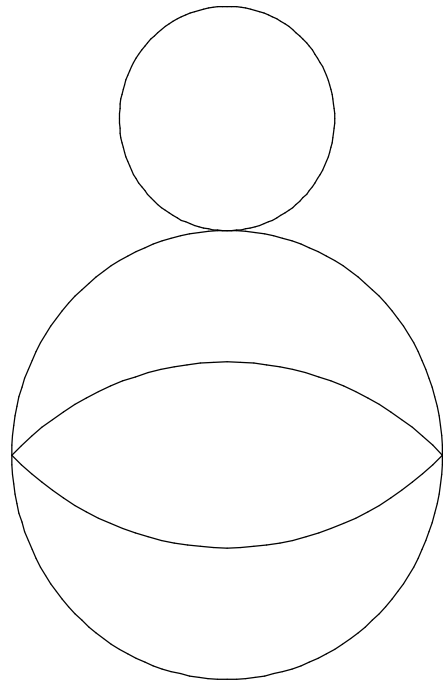
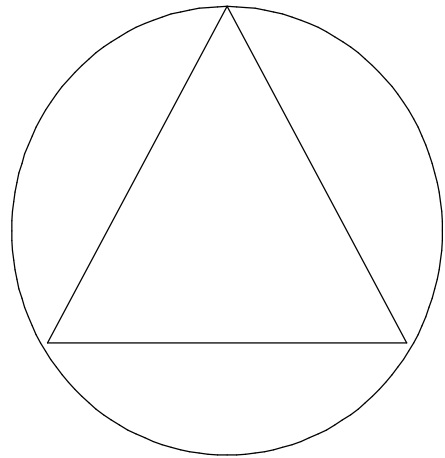
\end{document}